\documentclass[a4paper,11pt]{article}
\pdfoutput=1 

\usepackage{jcappub} 

\usepackage[T1]{fontenc} 
\usepackage{fontawesome5}
\usepackage{float}
\usepackage{soul}
\usepackage{amsmath}
\usepackage{subfigure} 
\usepackage{float}
\usepackage{subcaption}
\usepackage{multicol}
\usepackage{xcolor}

\newcommand{\ie}{{\em i.e.~}}
\newcommand{\eg}{{\em e.g.~}}

\title{\boldmath Gravitational Waves from Hyperbolic Encounters of Primordial Black Holes in Dwarf Galaxies}

\author[a]{Tadeo D. Gómez-Aguilar,}
\author[b]{Encieh Erfani,}
\author[c]{and N. M. Jiménez Cruz\,}

\affiliation[a]{Instituto de Ciencias Físicas, Universidad Nacional Autónoma de México, 62210, Cuernavaca, Morelos, México}
\affiliation[b]{Perimeter Institute for Theoretical Physics, Waterloo, ON N2L 2Y5, Canada}
\affiliation[c]{Physics Department, Swansea University, SA28PP, United Kingdom}

\emailAdd{tadeo.dga@icf.unam.mx}
\emailAdd{eerfani@perimeterinstitute.ca}
\emailAdd{2330189@swansea.ac.uk}

\abstract{We investigate the stochastic gravitational wave background (SGWB) generated by primordial black holes (PBHs) in the dense cores of dwarf galaxies (DGs), considering both hierarchical binary black hole (BBH) mergers and close hyperbolic encounters (CHEs). Extending our previous merger framework, we incorporate up to four successive generations of PBHs within a Hubble time and quantify the GW emission from both channels. Our results show that while BBHs dominate the total emission, CHEs occur earlier, provide the first GW signals, and contribute a continuous though subdominant background that becomes relatively more significant once the initial PBH population is depleted and binary formation is suppressed. We compute the resulting SGWB spectra, demonstrating that BBHs and CHEs imprint distinct frequency dependencies consistent with analytical expectations. We then compare the predicted signals with the sensitivity of observatories such as LISA, DECIGO, ET, IPTA, and SKA. The numerical implementation is publicly available at \href{https://github.com/TadeoDGAguilar/PBHs_and_GWs_into_DG}{\faGithubSquare}~\texttt{HierarchicalCHEs}.}

\begin{document}
\maketitle
\flushbottom

\section{Introduction}
\paragraph{}
Among many candidates proposed to account for the dark matter (DM) component of the Universe, primordial black holes (PBHs) remain as one of the intriguing possibilities \cite{Carr:2016drx, ESCRIVA2024261}. Formed in the early Universe from the gravitational collapse of large density fluctuations, PBHs could span a wide range of masses \cite{Zeldovich:1967lct, Hawking:1971ei, Carr:1974nx}. 
If PBHs survive to the present day due to Hawking radiation \cite{Hawking:1975vcx}, they could be compelling DM candidates \cite{carr2022primordial, green2021primordial}.
One auspicious environment to explore these PBHs is dwarf galaxies (DGs). Their small stellar populations, low metallicity, and shallow potential wells make them ideal laboratories to investigate the clustering, dynamics, and potential observational signatures of PBHs as DM candidates. Recent studies have shown that the internal kinematics and density profiles of DGs can accommodate the presence of dense PBH clusters without violating observational constraints \cite{Zhu:2017plg, Stegmann:2019wyz}.

There are several methods for the detection of DM PBHs (For a detailed review, see \cite{Carr:2020gox, Carr:2023tpt}). However, the detection of gravitational waves (GWs) by the LIGO/Virgo \cite{LIGOScientific:2016aoc} has opened new avenues for probing the origin and evolution of black holes (BHs) \cite{Bird:2016dcv, Clesse:2016ajp}. 
PBHs, if they exist and survive the Hawking radiation, will undergo gravitational interactions both among themselves and with astrophysical BHs. Such interactions can, in principle, generate a stochastic gravitational wave background (SGWB) through binary formation and dynamical interactions within the detectable range of proposed GW detectors, depending on the PBH mass and abundance \cite{LISACosmologyWorkingGroup:2023njw}.

In our previous work \cite{Erfani:2022gno}, we studied the GWs from binary formation of PBHs in the core of DGs. The core of these galaxies are dense enough for hierarchical merger of PBHs since it prevents merger remnants from being ejected by GW recoils\footnote{After our work, Ref.~\cite{Siles:2024yym} studied PBH mergers in clusters using N-body simulations, showing that GW recoil from asymmetric-mass, spinning binaries can suppress hierarchical mergers. In \cite{Erfani:2022gno}, we considered spinless PBHs, for which recoil velocities are reduced, and hierarchical mergers are therefore not excluded within this modeling assumption.}. We adopted a Plummer type density profile \cite{10.1093/mnras/71.5.460} and assumed a monochromatic PBH population constituting all of the DM. By considering the dynamical interactions within the core of these galaxies, we evaluated the formation of gravitationally bound binaries and tracked their hierarchical mergers over cosmic time. This approach provides a physically consistent and observationally motivated framework to explore the GW signatures of PBHs in dense environments such as DGs. We found that up to four successive merger generations can occur within a Hubble time. While the initial population was monochromatic, the cumulative effect of mergers introduced a spread in the mass spectrum. Our significant result was that both the total mass loss in the DG core due to GW emission and the fraction of BHs involved in collisions are largely independent of the initial PBH mass.
 
In addition to binary formation and mergers of PBHs in dense environments, another source of a SGWB can arise due to Close Hyperbolic Encounters (CHEs) \cite{Garcia-Bellido:2017qal}, in which two PBHs undergo a strong gravitational interaction without forming a bound system. These unbounded interacting PBHs can emit gravitational radiation, especially in environments where large number densities and moderate velocity dispersions favor high encounter rates. While the GW emission from binary mergers has been extensively studied, recent work has also investigated the contribution of CHEs to the SGWB \cite{Garcia-Bellido:2021jlq}. Motivated by this, we extend our previous work to incorporate the evaluation of CHEs rates and their associated gravitational radiation. This work aims to quantify the role of CHEs in the dynamical evolution of PBH clusters in the dense core of DGs and assess their contribution to the overall GW signal expected from such environments. We estimate the SGWB amplitude arising from CHEs in DGs for the first time and examine its potential detectability in future GW experiments, comparing it with the contribution from binary black holes (BBHs).

The paper is organized as follows: In Section~\ref{hierarchical_merger}, we review the hierarchical merger framework for PBHs in DG cores. In Section~\ref{CHE}, we develop the formalism for CHEs, deriving the encounter dynamics, cross sections, and associated GW emission. This provides a consistent extension of the merger analysis, allowing for a direct comparison between the GW backgrounds from BBHs and CHEs. In Section~\ref{Results}, which is devoted to our results, we present the numerical setup, parameter scans, and main results, including the time/redshift evolution of BBHs and CHEs, as well as their contribution to the GW emission. Finally, Section~\ref{conclusions} summarizes our findings and outlines future directions.
\section{Hierarchical Mergers of Primordial Black Holes in Dwarf Galaxies}\label{hierarchical_merger}
\paragraph{} 
In our previous work~\cite{Erfani:2022gno}, we investigated hierarchical PBH mergers in a scenario where PBHs constitute the dominant dark matter (DM) component in the cores of typical dwarf galaxies (DGs)\footnote{These galaxies are natural environments for PBHs due to several reasons. For more details, see \cite{Clesse:2017bsw}.}.
We evaluated binary PBH (BBH) formation, merger rates, and GW emissions from total mass loss across successive generations.
The framework established in that work permits an extension of the analysis to gravitational wave emission arising from PBH mergers and binary formation. In this section, we review those results, and then in the subsequent section, we consider the contribution from PBHs that do not form binaries but instead undergo close hyperbolic encounters (CHEs).

We considered a DG with mass \( M_{\rm DG} = 10^{9} M_{\odot} \) and radius \( R_{\rm DG} \sim 10\) pc, with a dense cluster of monochromatic PBH in its center of \( M_{\rm c} = 10^{5} M_{\odot} \) and \( R_{\rm c} \approx 1 \) pc, consistent with observations~\cite{Simon:2007dq}. We investigate a population of PBHs situated at the core of the DG, comprising $N_{\rm PBH}$ objects, which we refer to as a \textit{cluster}. We also assumed that PBHs account for the entirety of the DM content and that they constitute a monochromatic population characterized by a single initial mass, $m_{\rm PBH}$. Therefore, the number of PBHs in the core is
\begin{equation}\label{number_PBH}
N_{\rm PBH} = \frac{4\pi}{3} \frac{R_{\rm c}^3}{m_{_{\rm PBH}}}\,.
\end{equation} 
The four different initial masses for PBHs, which we define as \textit{first generation (1G)}, and their number at the core are reported in Table \ref{Tab:Initial_masses}.
\begin{table}[H]
\centering
\begin{tabular}{|c|c|c|c|c|}
\hline
\( m_{_{\rm PBH}} \) & \( 10^{-14}\,M_{\odot} \) & \( 10^{-2}\,M_{\odot} \) & \( 1\,M_{\odot} \) & \( 10\,M_{\odot} \) \\ \hline
\( N_{\rm PBH}\) & \( 10^{19} \) & \( 10^{7} \) & \( 10^{5} \) & \( 10^{4} \)\\ \hline
\end{tabular}
\caption{Initial (1G) masses of PBHs and their abundance in the core of a DG.}
\label{Tab:Initial_masses}
\end{table}
The merger starts with the involvement of equal-mass 1G PBHs; \ie (\(1\mathrm{G}+1\mathrm{G}\)) which produces second-generation (2G) BHs with roughly double the mass. Subsequent mergers (\(1\mathrm{G}+2\mathrm{G}\), \(2\mathrm{G}+2\mathrm{G}\), \(1\mathrm{G}+3\mathrm{G}\), \(2\mathrm{G}+3\mathrm{G}\), \(3\mathrm{G}+3\mathrm{G}\)) yield a variety of masses (See Table~\ref{Tab:cluster_evolution}).

The spatial distribution of PBHs at the core follows a Plummer density profile\footnote{We use the Plummer model since it is a widely used analytic model in astrophysics to describe the density distribution and gravitational potential of spherical stellar systems or DM clusters, such as globular clusters, and galactic bulges.}\cite{10.1093/mnras/71.5.460}. The density and the gravitational potential are
\begin{equation}
\rho_{i}(r) = \frac{3\,m_{i}\,N_{i}}{4\pi\,R_{\rm c}^{3}}\left( 1 + \frac{r^{2}}{R_{\rm c}^{2}}\right)^{-5/2}\,, \qquad
\phi(r) = \frac{GM_{\rm c}}{R_{\rm c}}\left( 1 + \frac{r^{2}}{R_{\rm c}^{2}}\right)^{-1/2}\,,
\label{eq:profile}
\end{equation}
where $i$ stands for the different species of PBHs.\\
The cross-section for gravitational capture and binary formation of two objects with masses \(m_i\) and \(m_j\) is given by \cite{1989ApJ...343..725Q, Mouri:2002mc}  
\begin{equation}\label{eq:sigma_BBH}
\sigma_{\rm BBH}(m_i,\,m_j) = 2 \pi \left(\frac{85\pi}{6\sqrt{2}}\right)^{2/7}\frac{G^{2}(m_{i} + m_{j})^{10/7}(m_{i}m_{j})^{2/7}}{c^{10/7}\,{|v_i - v_j|^{18/7}}}\,. 
\end{equation}
We model the root mean-squared (rms) velocity of each PBH in the \textit{cluster} as
\begin{equation}\label{rms velocity}
\overline{v_{i}^{2}}(r) = \frac{4\pi\,m_{i}}{\rho_{i}(r)}\int_{0}^{\phi(r)} f_{i}(E) \left(2(\phi(r)-E)\right)^{3/2}\,dE\,,
\end{equation}
where the distribution function of each PBH population with energy $E_i = \frac{G\,m_i^2N_i^2}{2R_{\rm c}}$ is \cite{1989ApJ...343..725Q} 
\begin{equation}
f_{i}(E) = \frac{96\sqrt{2}}{28\pi^3}\frac{R_{\rm c}^2}{G^5\,M_{\rm c}^5}\,E_i^{7/2}\,.
\label{eq:distribution_func}
\end{equation}

According to the Fokker-Planck equation that accounts for the loss and gain of BHs due to mergers with other BHs, the merger rate is given by \cite{stasenko2021merger, 1989ApJ...343..725Q}
\begin{equation}\label{eq:BBHmergerRate}
\Gamma_{j}(r) = \frac{14\pi}{3}\sum_{i}\,\sigma_{\rm BBH}(m_i,\, m_j)\int dr \, r^2 \left( \frac{n_{i}}{\overline{v_i}} \right)\left(\frac{n_j}{\overline{v_j}}\right) \left[(\overline{v_i} + \overline{v_j})^{3/7} - |\overline{v_i} - \overline{v_j}|^{3/7}\right]\,,
\end{equation}
where $n_{i}(r)$ is the number density of PBHs at radius $r$.\\
Having determined the merger rate, we can calculate the corresponding energy loss due to gravitational radiation via the well-known time-averaged energy loss rate in Keplerian orbits \cite{Celoria:2018mzr, Peters:1963ux, Sasaki:2018dmp}
\begin{equation}
\left\langle\frac{dE}{dt}\right\rangle = -\frac{32}{5}\frac{G^4(m_im_j)^2(m_i+m_j)}{a^5} F(e)\,, 
\end{equation}
where $e$ and $a$ are the eccentricity and semi-major axis of the orbit, respectively, and the explicit dependence on the eccentricity is given by
\begin{equation}
F(e) = \frac{1}{(1-e^2)^{7/2}}\left( 1+\frac{73}{24}e^2 + \frac{37}{96}e^4\right)\,.
\end{equation}
It is straightforward to show that the energy loss in GWs during binary formation can be expressed as \cite{Erfani:2022gno}
\begin{equation}\label{eq:EnergyEmission_BBH}
\Delta E_{\rm BBH} = \frac{1}{2}m_i m_j \left(\frac{1}{a_{\rm merge}} - \frac{1}{a_{\rm i}}\right)\,,
\end{equation}
where $a_{\rm i}$ and $a_{\rm merge}$ are the initial separation and sum of the Schwarzschild radii of the progenitors, respectively.

We found that, in dense DG cores, up to four successive merger generations, \ie, four epochs \cite{Erfani:2022gno} can take place due to the age of the Universe. Table~\ref{Tab:cluster_evolution} presents the resulting BH masses and their abundances for an initial (1G) mass of $1\,M_{\odot}$. At each epoch (redshift interval), the BHs formed through hierarchical mergers are indicated in bold.
\begin{table}[h!]
\centering
\begin{tabular}{|c|c|c|c|c|c|c|c|c|c|c|c|}
\hline
\multicolumn{12}{|c|}{\textbf{First Epoch}\,($z = 20 - 1.88$)} \\ \hline
$m_i$ & 10 & \multicolumn{10}{c|}{}\\ \hline
$N_i$ & $10^4$ & \multicolumn{10}{c|}{} \\ \hline
$v_i$ & $\sim 13$ & \multicolumn{10}{c|}{} \\ \hline
\multicolumn{12}{|c|}{\textbf{Second Epoch}\,($z = 1.88 - 1.02$)} \\ \hline
$m_i$ & 10 & \textbf{18.7} & \multicolumn{9}{c|}{}  \\ \hline
$N_i$ & 6434 & 1783 & \multicolumn{9}{c|}{}  \\ \hline
$v_i$ & $\sim 10.6$ & $\sim 7.5$ & \multicolumn{9}{c|}{} \\ \hline
\multicolumn{12}{|c|}{\textbf{Third Epoch}\,($z = 1.02 - 0.44$)} \\ \hline
$m_i$ & 10 & 18.7 & \textbf{27.1} & \textbf{35.2} & \multicolumn{7}{c|}{}\\ \hline
$N_i$ & 4955 & 994 & 1092 & 43 & \multicolumn{7}{c|}{} \\ \hline
$v_i$ & $\sim 9$ & $\sim 5.6$& $\sim 7$ &  $\sim 1.5$ & \multicolumn{7}{c|}{} \\ \hline
\multicolumn{12}{|c|}{\textbf{Fourth Epoch}\,($z = 0.44 - 0.0$)} \\ \hline
$m_i$ & 10 & 18.7 & 27.1 & 35.2 & \textbf{35.3} & \textbf{43.1} & \textbf{43.3} & \textbf{50.8} & \textbf{51.0} & \textbf{58.6} & \textbf{66.2}  \\ \hline
$N_i$ & 3154 & 266 & 526 & 12 & $833$ & $281$& 19 & 12 & 17 & 7 & $\sim 0$  \\ \hline
$v_i$ & $\sim 7.5$ & $\sim 2.8$ & $\sim 4.8$ & $\sim 1$ & $\sim 6.8$ & $\sim 4.4$ & $\sim 1.1$ & $\sim 1$ & $\sim 1.1$ & $\sim 0.77$ & - \\ \hline
\end{tabular}
\caption{BH masses (in $M_{\odot}$), velocities (in km/s), and abundances across successive merger generations in a dense DG core, assuming an initial 1G mass of $10\,M_{\odot}$. BHs formed via hierarchical mergers at each redshift $z$ are shown in bold.}
\label{Tab:cluster_evolution}
\end{table}
Note that while the initial population is monochromatic, the cumulative effect of mergers introduced a spread in the mass spectrum. At the end of each epoch, due to the formation of new BHs, the distribution of mass in the \textit{cluster}, and in consequence, the initial conditions for the next epoch change. 

In the following section, we extend our previous analysis to investigate alternative dynamical channels for GW emission in the core of DGs through close hyperbolic encounters (CHEs).
\section{Gravitational Waves from Close Hyperbolic Encounters of PBHs}\label{CHE}
\begin{figure}[h!]
\centering
\includegraphics[width=0.6\linewidth]{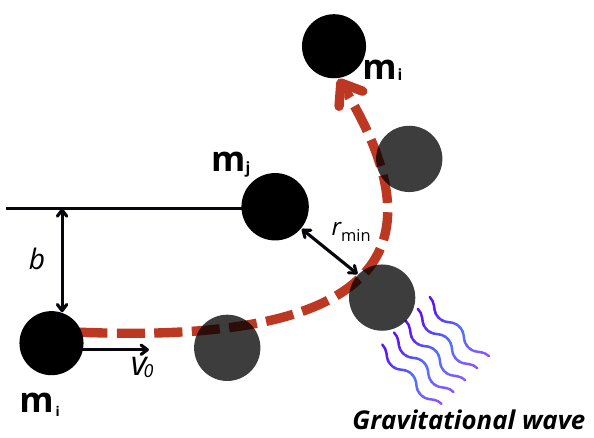}
\caption{Illustration of two BHs with masses $m_i$ and $m_j$ with relative velocity, $v_0$, in a close hyperbolic encounter, where $r_{\rm min}$ is the closest approach to produce GWs in this process.}
\label{Image1}
\end{figure}

\paragraph{}
In the dense environment of DG cores, most PBH encounters do not lead to the formation of bound systems but instead result in scattering events. During such gravitational interactions, PBHs may also undergo close hyperbolic encounters (CHEs), in which they emit GWs and thereby contribute to the stochastic gravitational wave background (SGWB) (see Fig.~\ref{Image1}).
The radiation emitted results from the rapid variation of the quadrupole moment during the closest approach of two BHs \cite{Garcia-Bellido:2017knh}. The encounter can be modeled in a non-relativistic regime as a Keplerian hyperbolic orbit characterized by the total mass $m_i + m_j$, relative velocity $v_0 = |v_i -v_j|$, and impact parameter $b = G\,(m_i + m_j)\sqrt{e^2-1}/{v_0^2}$. A key parameter in hyperbolic motion is the eccentricity, which is given by \cite{Caldarola:2023ipo}
\begin{equation}
e = \sqrt{1 + \frac{b^2 v_0^4}{G^2 (m_i + m_j)^2}}\,.
\end{equation}

To evaluate the GW energy loss, we first compute the scattering cross-section governing CHE dynamics, following an approach analogous to that used for binary formation. For a pair of unbound BHs with masses $m_i$ and $m_j$ and relative velocity $v_0$, the gravitational scattering cross-section is given by \cite{Mouri:2002mc}
\begin{equation}\label{eq:sigma_CHE}
\sigma_{\rm CHE} (m_i,\,m_j) = \pi b^2 = \pi \left(\frac{G\,(m_i + m_j)}{v_0^2}\right)^2 (e^2 - 1)\,.
\end{equation}
Thus, the total energy radiated in GWs for two masses at the closest approach or periastron distance, $r_{\rm min} = GM (e - 1)/ v_0^2 $ is given by \cite{Caldarola:2023ipo}
\begin{equation}\label{eq:EnergyEmmited_CHE}
\Delta E_{\rm CHE} = - \frac{8}{15}\frac{G^{7/2}}{c^5} \frac{(m_i + m_j)^{1/2} m_i^2 m_j^2}{r_{\rm{min}}^{7/2}}f(e)\,,
\end{equation}
where
\begin{equation}\label{eq:f_excentric}
f(e) \equiv \frac{1}{(1+e)^{7/2}} \left[ 24 \cos^{-1}\left(-\frac{1}{e}\right)\left(1 + \frac{73}{24}e^2 + \frac{37}{96}e^4\right) + \sqrt{e^2 - 1} \left(\frac{301}{6} + \frac{673}{12}e^2\right)\right]\,.
\end{equation}
Note that to calculate the total amount of energy radiated in GWs in the \textit{cluster} by CHEs, we divided the core into 10 shells as in \cite{Erfani:2022gno} and we multiplied the energy emitted for a pair of BHs, Eq.~(\ref{eq:EnergyEmmited_CHE}), by the total number of encounters in each shell. Summing over all shells and pair combinations yields the total GW emission from CHEs in the \textit{cluster}.

In the next section, we present our results for the GW emission from both BBH mergers and CHEs. We quantify the respective contributions of these two channels to the SGWB, highlighting the relative significance of hierarchical mergers and single scattering events in dense DG cores. The calculation of the SGWB generated by CHEs closely follows the formalism developed in \cite{Garcia-Bellido:2021jlq}, upon which our analysis builds with updated astrophysical inputs and rate estimates relevant for DGs environments.
\section{Results: SGWB from BBHs and CHEs}\label{Results}
\paragraph{}
Since the calculation of GW emission follows a similar approach for both BBHs and CHEs, we describe the simulation methodology in detail here, highlighting distinctions between the two cases where relevant, while noting that the underlying assumptions are the same. Note that although GW emission reduces the core mass, the effect is minor ($\sim 5\%$), allowing us to treat the core density as effectively constant.

We track the number of mergers that begin with a monochromatic population of PBH at redshift $z=20$. This choice ensures an initial population early enough to exclude stellar origins for the BHs, yet late enough to allow the formation of DGs. In the first iteration, the single-mass population is divided into two identical sets, treated as distinct species for input into our merging algorithm. The galaxy core is divided into 10 shells, with the number of mergers estimated under the assumption of a constant merger rate within each epoch. At every epoch, we update the initial conditions, evaluate the relevant cross-sections, and compute the corresponding GW energy loss.

The simulation was carried out using a custom Python code that directly implements the analytical formalism outlined in Sections~\ref{hierarchical_merger} and \ref{CHE}. The numerical code is public on the repository \href{https://github.com/TadeoDGAguilar/PBHs_and_GWs_into_DG}{\faGithubSquare}~\texttt{HierarchicalCHEs}. Note that in our simulations, we assume isotropic velocity distributions when computing merger rates and dynamical timescales. The current implementation does not include three-body interactions, tidal disruptions from external fields, or remnant ejection due to GW recoil.

We consider four groups of initial monochromatic populations of PBHs reported in Table \ref{Tab:Initial_masses}, all within the core of the \textit{cluster} described in Section~\ref{hierarchical_merger}. The total number of PBHs and their distribution are given by Eqs.~\eqref{number_PBH} and \eqref{eq:profile}. We construct all possible combinations of PBH pairs; \eg 1G -- 1G, 1G -- 2G (interactions of original population with the 2nd generation of PBHs resulting from the binary merger of original PBHs), 2G -- 2G, 1G -- 3G, and so on.
\begin{figure}[t!]
\centering
\subfigure[]{\includegraphics[scale=0.3]{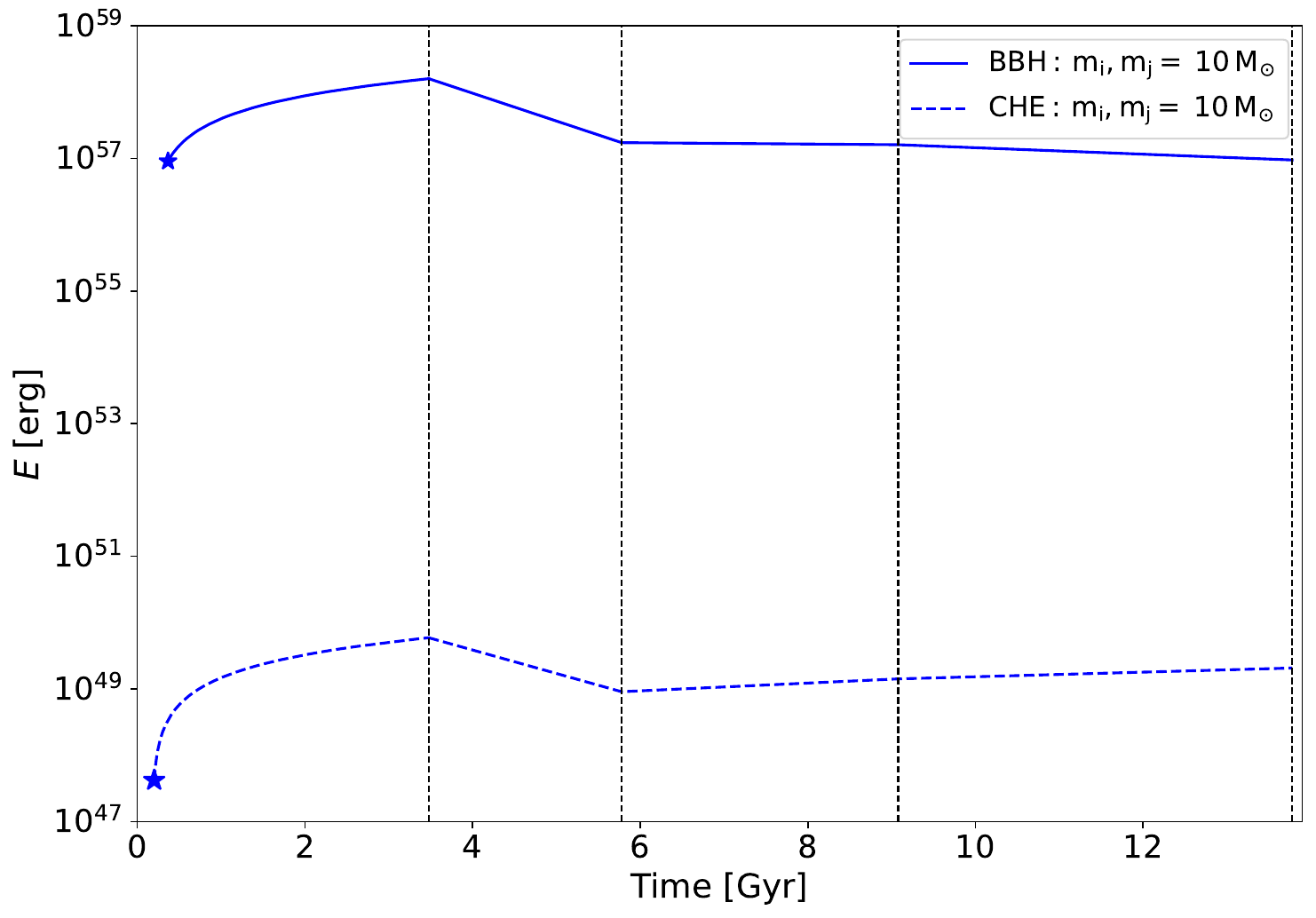}}
\hfill
\subfigure[]{\includegraphics[scale=0.3]{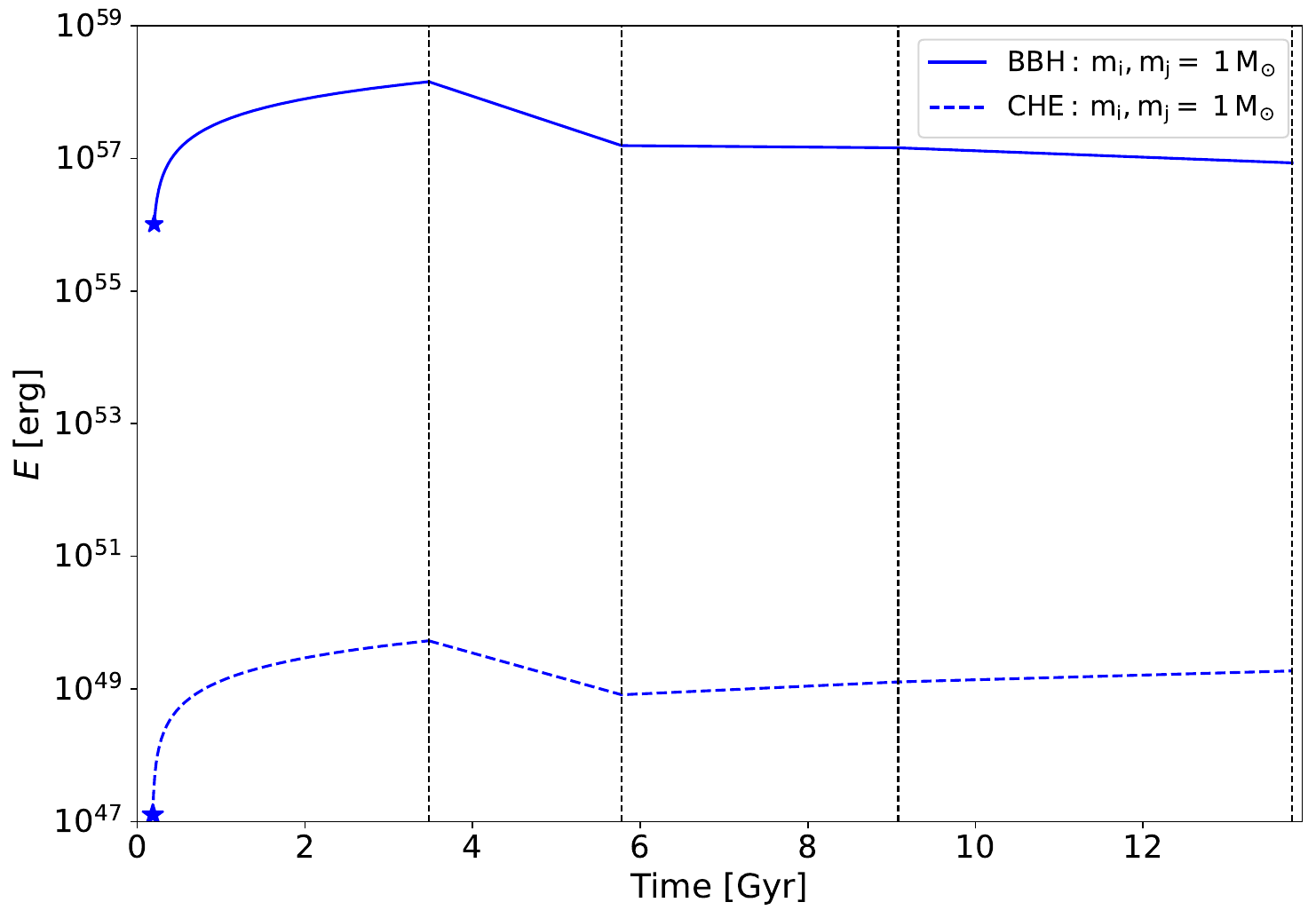}}
\subfigure[]{\includegraphics[scale=0.3]{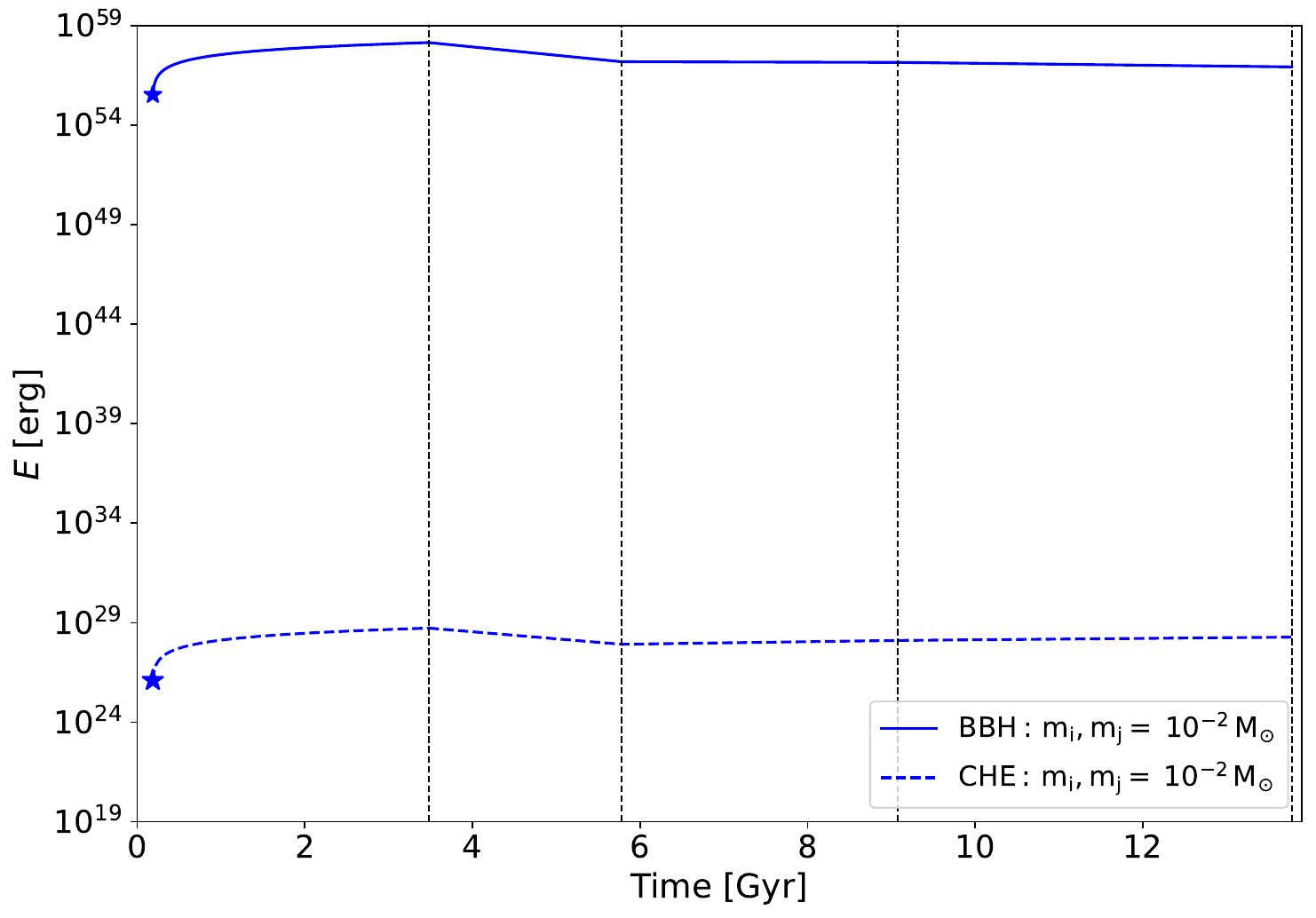}}
\hfill
\subfigure[]{\includegraphics[scale=0.3]{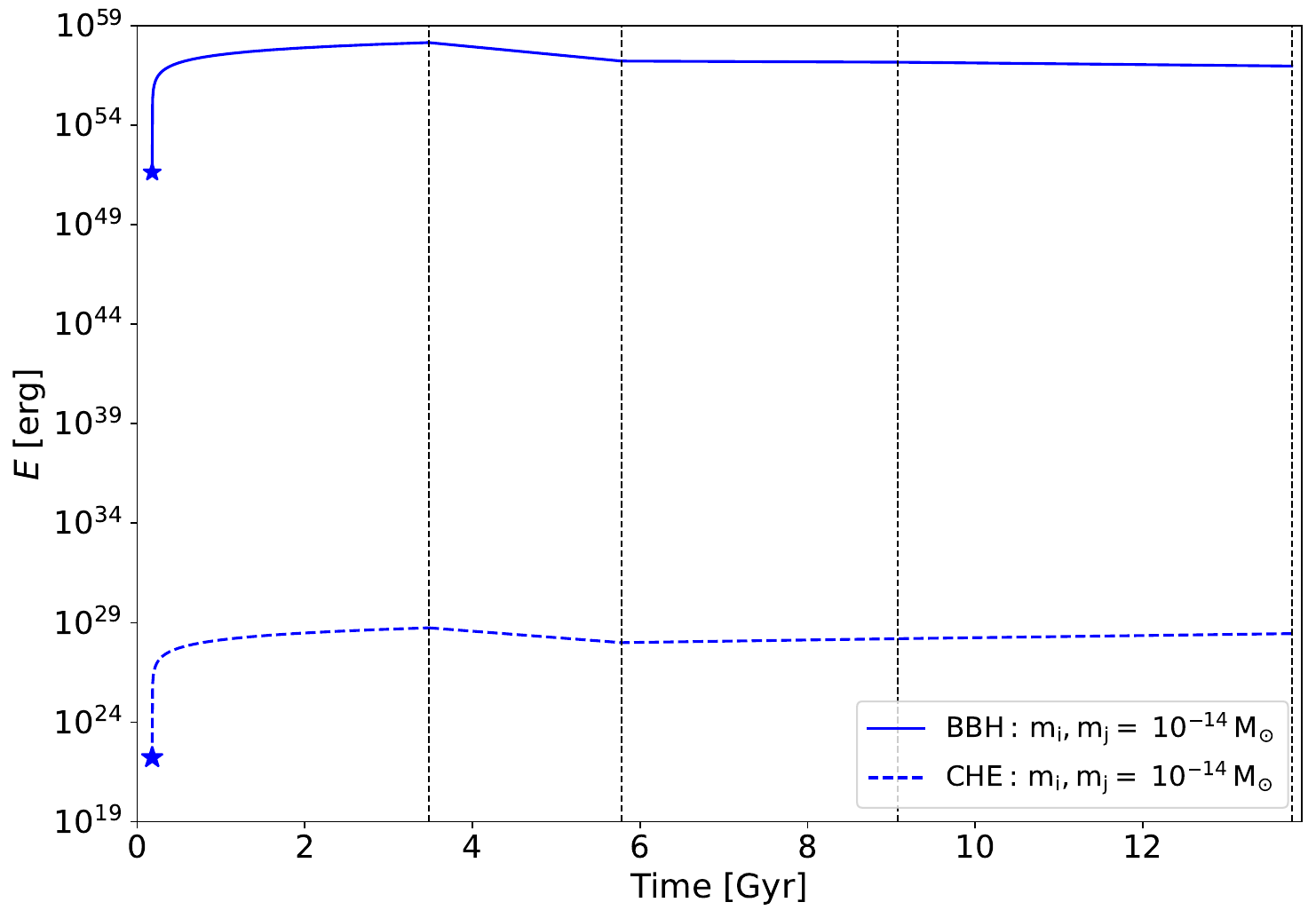}}
\caption{Energy loss from GW emission due to binary formation (solid line) and CHEs (dashed line) for the initial PBH population. Stars indicate the onset times of binary formation and CHEs. Vertical dotted lines mark successive epochs, with up to four BH generations forming within a Hubble time.}
\label{fig:1G_evolution}
\end{figure}
The gravitational potential (Eq.~\eqref{eq:profile}), rms velocity, and energy distribution function (Eq.~\eqref{eq:distribution_func}) are computed self-consistently from the adopted profile. To avoid the nonphysical divergence of the density as $r\to0$ (and the associated numerical instabilities when discretizing the Plummer profile), we employ a softening (or regularization) by introducing a small parameter $\delta_r$ such that $r \to r + \delta_r$, widely used in collisionless N‑body codes to stabilize central forces and densities~\cite{Athanassoula:1999wz, Power:2002sw}. Typical values of $\delta_r$ are chosen to be several orders of magnitude smaller than $R_{\rm c}$, sufficient to regularize the central divergence without altering global properties of the \textit{cluster} (See \cite{Erfani:2022gno} for more details.).

During the pre-evolutionary stage (\ie before the CHE/binary formation), the cluster remains in a quasi-stationary configuration. The initial probability for BBHs or CHEs is negligible, consistent with analytic estimates for time scales between hierarchical mergers. Thus, the softened profile establishes a stable baseline for all subsequent dynamical evolution\footnote{Adjustments to the regularization scale, initial assumptions, or the inclusion of further physics (\eg, three-body effects or GW recoil) may affect both the onset time and the resulting GW emission \cite{Gnedin:1998bp, Lousto:2011kp, Campanelli:2007cga, Varma:2022pld}.}.

Based on the original code implementation, we define fixed evaluation epochs to sample the system before the onset of dynamical activity. Within this initial phase, we find that the minimum timescale for CHEs to occur across all concentric shells of the core is shorter than that for the 1G PBH binary formation. For instance, with an initial mass of $10\,M_{\odot}$, the first CHEs appear at $\sim0.024\,{\rm Gyr}$, whereas the earliest binary mergers occur at $\sim0.18\,{\rm Gyr}$. These characteristic timescales are indicated by stars in Fig.~\ref{fig:1G_evolution}, where the solid (dashed) blue line denotes the cumulative GW emission from BBHs (CHEs) across all radii of the \textit{cluster}. After a binary formation, a time lapse of $\sim 0.38\,{\rm Gyr}$ is needed after the start of evolution to ensure that at least the first merger occurs which leads to the formation of the 2G of BHs population with mass of $\sim 18.7\,M_{\odot}$ (see Fig.~\ref{fig:1G_evolution}). This shorter time lapse motivates an extended exploration of the system with finer time resolution and additional evolutionary steps. 

The simulation results presented in Table~\ref{Tab:cluster_evolution} are directly implemented to construct Fig.~\ref{fig:1G_evolution}. In the first epoch ($z=20 - 1.88$), the system is dominated by $10\,M_{\odot}$ initial PBHs ($N_{\rm PBH}=10^4$) with velocities of $\sim 13$ km/s. During the second epoch ($z=1.88 - 1.02$), the first 2G BHs of mass $18.7\,M_{\odot}$ appear with abundance $N\simeq 1783$ and lower velocities ($\sim7.5$ km/s) compared to their progenitors. By the third epoch ($z=1.02 - 0.44$), heavier remnants emerge: $27.1\,M_{\odot}$ with $N\simeq1092$ and $35.2\,M_{\odot}$ with $N\simeq43$. In the final epoch ($z=0.44 - 0$), while the abundance of 1G PBHs has declined to $N\simeq3100$, the mass spectrum broadens, reaching $66.2\,M_{\odot}$, but with very low abundances ($N_i < 20$ for most of these high-mass species). This progression reveals two key trends: (i) lighter PBHs are gradually depleted over cosmic time, and (ii) heavier remnants emerge successively, but in significantly smaller numbers.

Fig.~\ref{fig:1G_evolution} also shows that the GW emission from CHEs is significantly lower than that from BBHs. Up to the formation of the 2G (first epoch), the energy loss from the initial PBH population increases for both channels, as 1G PBHs dominate the \textit{cluster} mass. After 2G formation, the energy loss from 1G binaries declines, reflecting the reduced 1G population. Notably, beyond the second epoch, the CHE contribution becomes slightly more pronounced relative to BBHs, indicating that while binary formation is suppressed as the 1G population decreases, CHEs continue to provide a non-negligible GW contribution.
\begin{figure}[ht!]
\centering
\includegraphics[width=1.\linewidth]{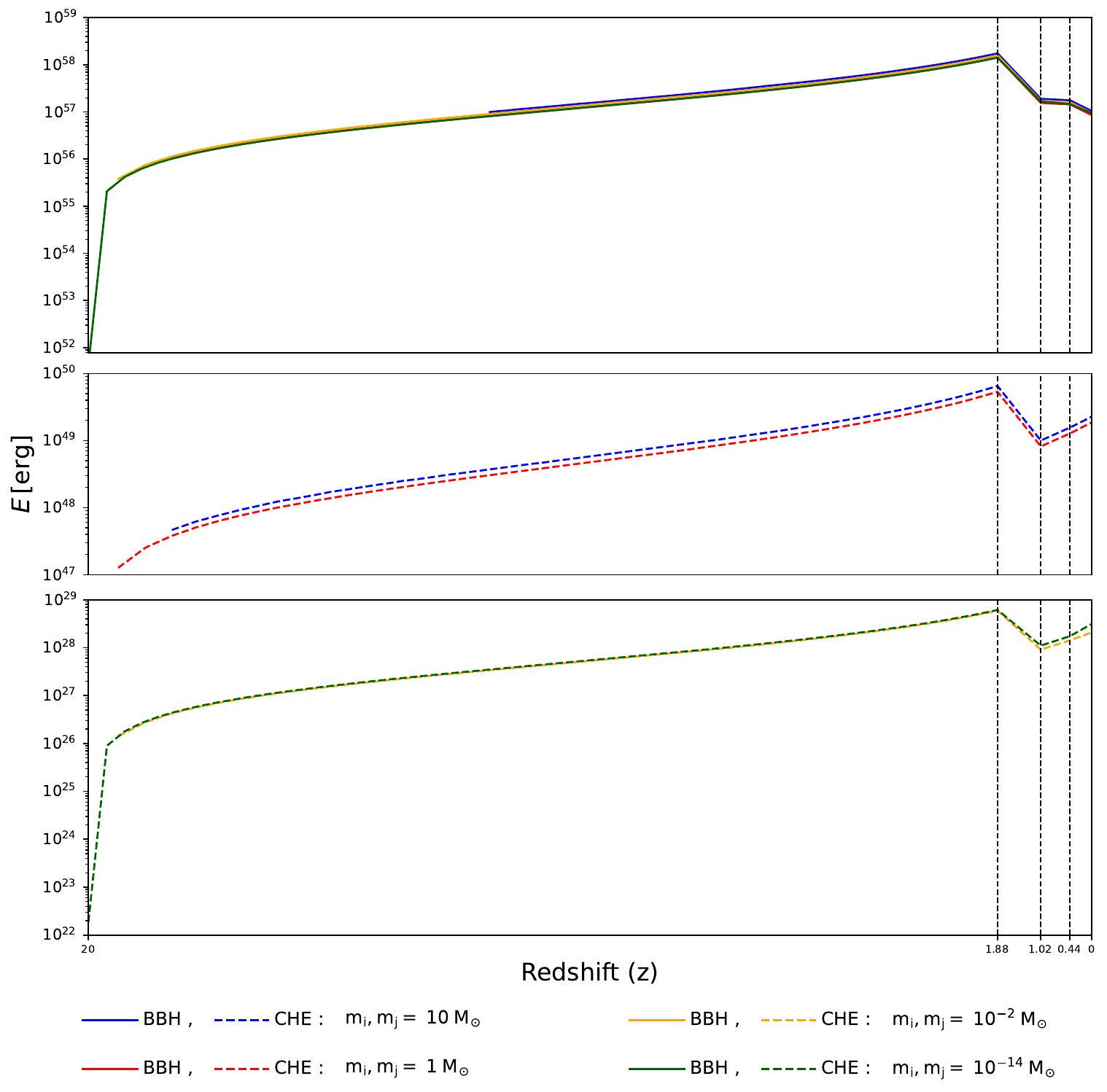}
\caption{Redshift evolution of GW emission from BBHs and CHEs for four initial masses. CHEs appear earlier than BBH formations, and while BBH contributions decrease at lower redshifts, CHEs continue to rise. Vertical dotted lines mark successive epochs in redshift.}
\label{fig:placeholder}
\end{figure}

The quantities described so far have been computed on a time grid. In Fig.~\ref{fig:placeholder}, we consider GW emission by BBHs and the CHEs for four different 1G masses wrt the redshift\footnote{For converting the time to redshift, we considered the flat $\Lambda$CDM model.}. This figure clearly illustrates that CHEs commence earlier than binary formations, and while GW emission from BBHs declines after the second epoch, it continues to increase for CHEs for the initial populations. 

\begin{figure}[ht!]
\centering
\includegraphics[scale=.60]{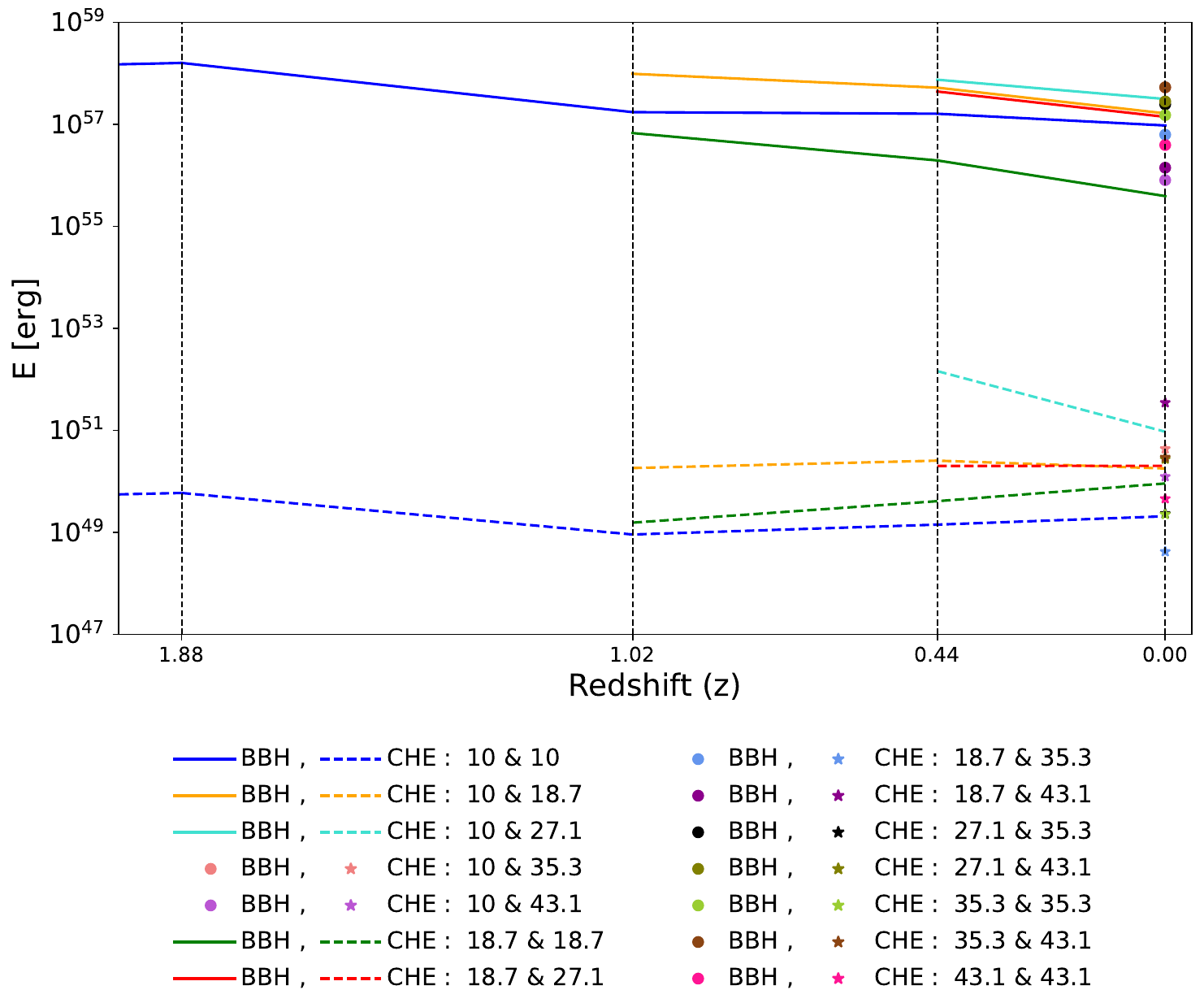}
\caption{GW emission from BBHs and CHEs for all generated masses, shown from the 2nd epoch ($z = 1.88$) to the present, for an initial PBH mass of $10\,M_{\odot}$.}
\label{fig:10M}
\end{figure}

\begin{figure}[ht!]
\centering
\includegraphics[scale=.65]{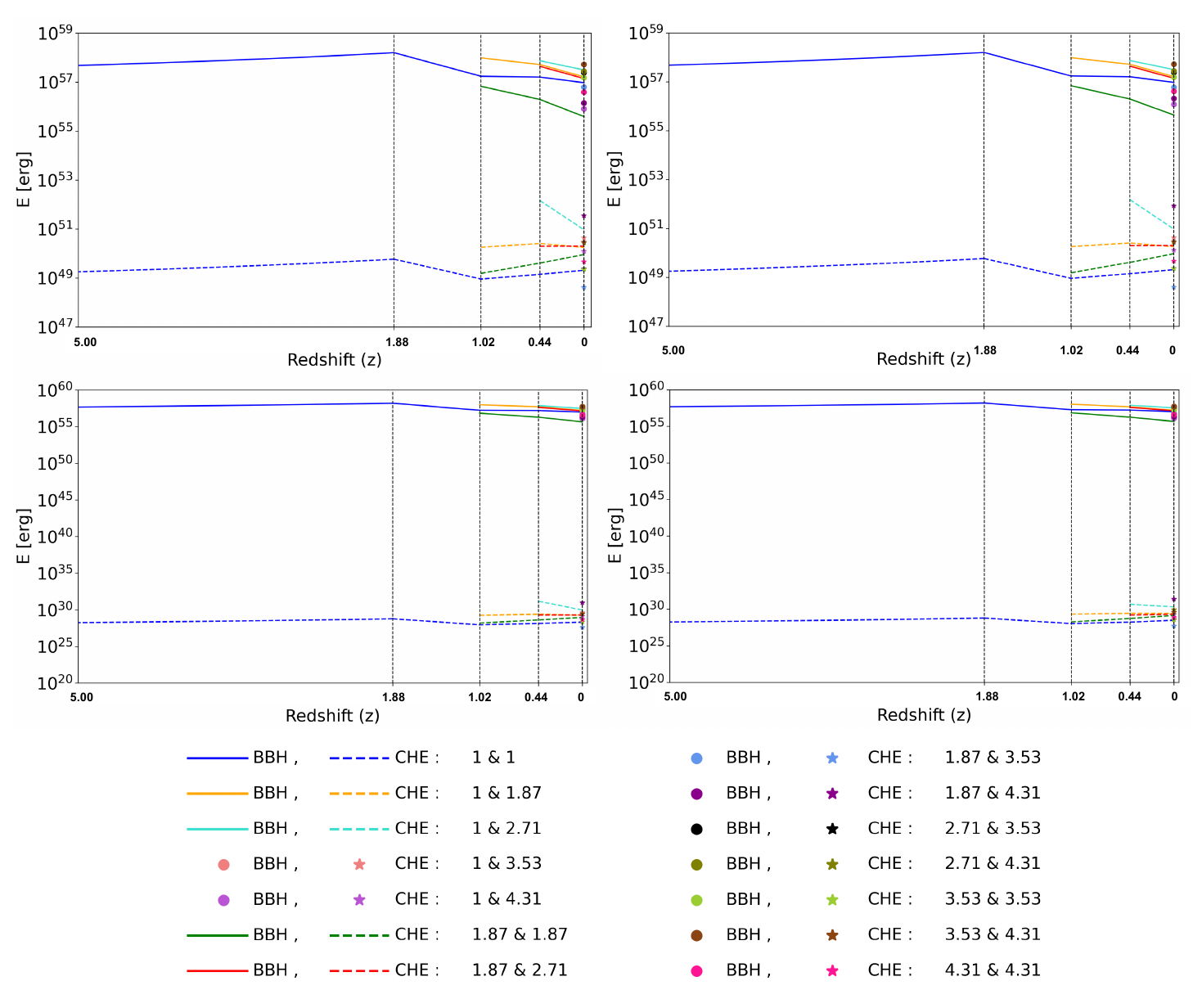}
\caption{GW emission from BBHs and CHEs across all PBH generations for four different initial masses. Results are shown from redshift 5 to the present, focusing on 2G PBHs and beyond. Upper left panel: values in the legend are multiplied by $10\,M_{\odot}$. Upper right panel: values in the legend are multiplied by $1\,M_{\odot}$. Lower left panel: values in the legend are multiplied by $10^{-2}\,M_{\odot}$. Lower right panel: values in the legend are multiplied by $10^{-14}\,M_{\odot}$.}
\label{fig:four_initial}
\end{figure}

The simulation structure enables detailed tracking of each PBH generation and interaction type, providing a foundation for estimating the SGWB for BBHs and CHEs. In Fig.~\ref{fig:four_initial}, we show the GW emission from two channels for all the generations of BHs for four different initial masses. Since our focus is on 2G BHs and higher, we present results from redshift 5 to the present. To better illustrate the results, Fig.~\ref{fig:10M} shows the GW emission for all generated masses from the 2nd epoch ($z = 1.88$) to the present, for an initial mass of $10\,M_{\odot}$.

With the GW emission from both channels established, we proceed to describe the resulting SGWB spectrum from BBHs and CHEs, followed by a comparison with current and future GW observations. The primary observable of this background is characterized by the relative GW energy density to the critical density $\rho_{\rm crit}$ today, per unit interval of logarithmic frequency in the frame of the observer \cite{Allen:1997ad}
\begin{equation}
\Omega_{\rm GW}(f) \equiv \frac{1}{\rho_{\rm crit}} \frac{d \rho_{\rm GW}}{d \ln{f}}\,.  
\end{equation}
For a SGWB of a point source origin, this dimensionless parameter can be written as \cite{2001astro.ph..8028P}
\begin{equation}\label{eq:omega_gw}
\Omega_{\rm GW}(f)=\dfrac{1}{\rho_{\rm crit}}\int_{0}^{\infty}dz\, \dfrac{\mathcal{R}(z)}{(1+z)^2 H(z)}\dfrac{dE_{\rm GW}}{d\ln f_{\rm ref}}\,.
\end{equation}
where $f_{\rm ref}=f(1+z)$ is the GW frequency in source frame, and $H(z)$ is the Hubble expansion rate. Note that the merger rate per unit time per comoving volume, $\mathcal{R}(z)$, and the energy emitted by a single source, $dE_{\rm GW}/d \ln f_{\rm r}$, are different for BBHs and CHEs. 

The associated SGWB is given by the following equations for the BBHs and CHEs, respectively (for more details see \cite{Garcia-Bellido:2021jlq}).
\begin{align}\label{eq:Omega_BBH}
\Omega_{\rm GW}^{\rm BBH}(f) \approx \, & \,2.39 \times 10^{-11}\,h_{70}\left(\frac{\Omega_{\rm DM}}{0.25}\right)^{2}\left(\frac{v_0}{10 \,{\rm km/s}}\right)^{-11/7}\left(\frac{\delta_{\rm c}}{10^{10}}\right)\left(\frac{f}{1\,{\rm Hz}}\right)^{2/3}\nonumber\\  
& \int dm_{i}dm_{j}\frac{f(m_i)f(m_j)(m_i + m_j)^{23/21}}{(m_im_j)^{5/7}}\,, 
\end{align} 
\begin{align}\label{eq:Omega_CHE}
\Omega_{\rm GW}^{\rm CHE}(f) \approx \, & \,9.81 \times 10^{-11} h_{70} \left(\frac{\Omega_{M}}{0.3}\right)^{-1/2}\left(\frac{\Omega_{\rm DM}}{0.25}\right)^{2} \left(\frac{\delta_{\rm c}}{10^{10}}\right) \left(\frac{a}{0.1\,\rm AU}\right) \left(\frac{y}{0.01}\right) \nonumber\\
& \left(\frac{f}{10 \,{\rm Hz}}\right)^{2} \int \frac{dm_{i}}{100\,M_{\odot}} \frac{dm_{j}}{100\,M_{\odot}} f(m_i) f(m_j) e^{-2 x_{0} \xi (y)} \tilde{I}[y,\,x_0]\,, 
\end{align}
where 
\begin{align}
x_0 & = 2\pi  \sqrt{\dfrac{a^3}{G (m_{i} +m_{j})}}\, f\,,\nonumber\\
\xi (y) & = y -\tan^{-1} y\,,\nonumber\\  
y & = \sqrt{e^2-1}\,,\nonumber\\
\tilde{I}[y,\,x_0] & = \dfrac{1-y^2+4y^2+\frac{3}{2}\frac{x_0 y^6}{\xi(y)}}{(1+y^2)^2}\,,\nonumber\\
\end{align}
and $\delta_{\rm c}$ is the local density contrast, and values $\sim 10^9 - 10^{10}$ are typical of the DM density in DGs detected by Keck/DEIMOS \cite{Martin:2007ic}. 

\begin{figure}
\centering
\includegraphics[width=1\linewidth]{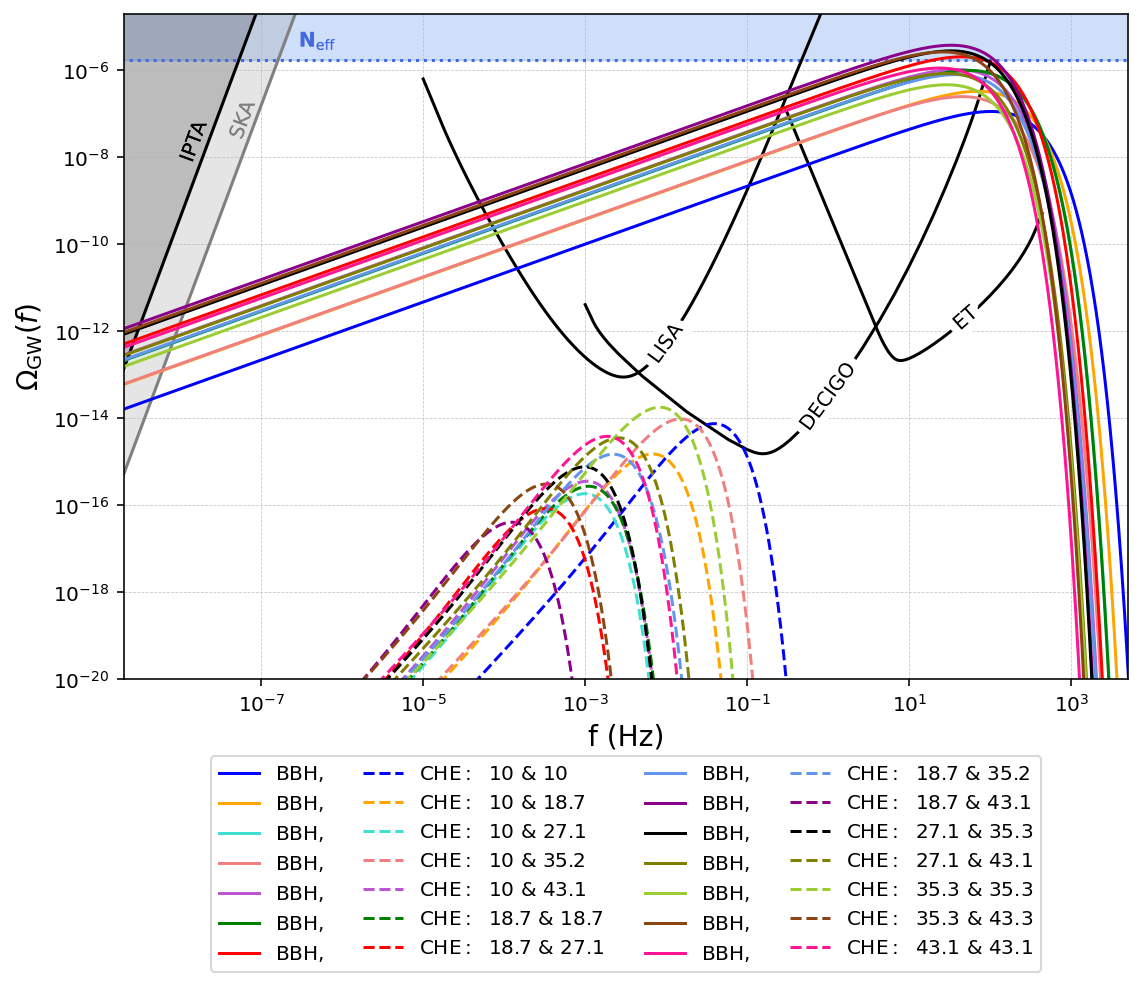}
\caption{Comparison of the SGWB spectra from BBHs (solid) and CHEs (dashed) for different 4G PBH mass combinations, assuming an initial mass of $10\,M_{\odot}$. The sensitivity curves for IPTA \cite{Hobbs:2009yy}, SKA \cite{Carilli:2004nx, Janssen:2014dka}, LISA \cite{Flauger:2020qyi}, DECIGO \cite{Kawamura:2011zz}, and ET \cite{Abac:2025saz} have been shown. The blue dotted line is the constraint from the effective number of neutrinos \cite{Caprini:2018mtu}, and this limit (blue shaded region) must not be exceeded.}
\label{fig:GW_spectrum}
\end{figure}

Using the GW energy results from Sections~\ref{hierarchical_merger} and \ref{CHE}, we present the GW spectra in Fig.~\ref{fig:GW_spectrum}. As lower initial masses yield weaker emission, we focus on the case of $10\,M_{\odot}$. Only the 4th epoch results for PBHs with $N_i > 100$, as listed in Table~\ref{Tab:cluster_evolution}, are considered. The immediate observation from Fig.~\ref{fig:GW_spectrum} is that the SGWB spectrum from CHEs exhibits a steeper slope than that from BBHs, consistent with the frequency dependence predicted by Eqs.~\eqref{eq:Omega_BBH} and \eqref{eq:Omega_CHE}. Furthermore, the BBH contribution is expected to be detectable by the Einstein Telescope (ET), while the CHE spectrum may reach the sensitivity of DECIGO. Although the CHE contribution remains below that of BBHs, it should be noted that our analysis does not account for binary disruption in dense environments, which could further reduce the GW amplitude.

\section{Conclusions}\label{conclusions}
\paragraph{}
In this work, we have investigated the stochastic gravitational wave background (SGWB) generated by primordial black holes (PBHs) in the dense cores of dwarf galaxies (DGs), extending our previous analysis of hierarchical PBH mergers to include the contribution from close hyperbolic encounters (CHEs). By simultaneously modeling both binary black hole (BBH) mergers and CHEs within the same dynamical framework, we provide a more comprehensive description of the GW signatures expected from clustered PBHs. The hierarchical nature of the system leads to a gradual depletion of light PBHs and the emergence of heavier remnants, with up to four successive merger generations possible within a Hubble time, distributed across four distinct epochs. This hierarchical evolution broadens the PBH mass spectrum and modifies the temporal distribution of GW emission in a manner largely independent of the initial PBH mass (see Fig.~\ref{fig:four_initial}).

Fig.~\ref{fig:10M} demonstrates that BBHs and CHEs contribute to the SGWB in complementary ways. Binary mergers dominate the total GW emission across cosmic time, particularly after the formation of second-generation PBHs, while CHEs occur earlier and represent the first source of GW production before the onset of binary PBH mergers. Although the cumulative energy radiated by CHEs remains below that of BBHs, they provide a continuous GW background across all epochs and become relatively more significant once the initial PBH population is depleted and binary formation is suppressed. The resulting SGWB spectra shown in Fig.~\ref{fig:GW_spectrum} indicate that the BBH contribution is dominant but that CHEs imprint a distinct spectral slope consistent with analytical expectations. Our analysis shows that CHEs, despite being subdominant in amplitude, may still become relevant for future GW observatories, particularly in frequency ranges probed by DECIGO. Their earlier onset compared to mergers also highlights their role as a probe of the early dynamical state of PBHs.

Several simplifying assumptions were made to render the problem tractable. We adopted a monochromatic initial mass function, spherical symmetry with a Plummer density profile, considering PBHs as only DM contribution, and a uniform overdensity region parametrized by $\delta_{\rm c}$, while neglecting N-body effects such as three-body interactions, tidal disruptions, and binary ejections due to GW recoil. These effects, along with more realistic PBH mass spectra and anisotropic velocity distributions, are expected to modify the merger and encounter rates and should therefore be included in future refinements of the model. 

\acknowledgments

We are grateful to Juan Carlos Hidalgo and Guillem Dom\`enech for insightful comments on the manuscript. EE is supported by the IIE-Scholar Rescue Fund, Canadian Institute for Theoretical Astrophysics (CITA), and the Perimeter Institute for Theoretical Physics. Research at Perimeter Institute is supported in part by the Government of Canada through the Department of Innovation, Science and Economic Development and by the Province of Ontario through the Ministry of Colleges and Universities. NMJC thanks the Science and Technology Facilities Council (STFC) for her research scholarship. The work of TDGA is supported by DGAPA-PAPIIT-UNAM grant No. IN110325 Estudios en cosmología inflacionaria, agujeros negros primordiales y energía oscura and SECIHTI grant CBF 2023-2024-162.

\bibliographystyle{ieeetr}
\bibliography{biblio}
\end{document}